\documentclass[conference]{IEEEtran}
\usepackage{graphicx}
\usepackage{amsmath}
\usepackage{multirow}
\usepackage{xspace}
\usepackage{adjustbox}
\usepackage[utf8]{inputenc}
\usepackage{hyperref}
\usepackage{array}
\newcolumntype{P}[1]{>{\centering\arraybackslash}p{#1}}
\newcolumntype{M}[1]{>{\centering\arraybackslash}m{#1}}
\usepackage{enumitem}

\newcommand*{\ep}{$E$\xspace}

\usepackage[underline=false]{pgf-umlsd}
\renewcommand{\mess}[4][0]{
	\stepcounter{seqlevel}
	\path
	(#2)+(0,-\theseqlevel*\unitfactor-0.7*\unitfactor) node (mess from) {};
	\addtocounter{seqlevel}{#1}
	\path
	(#4)+(0,-\theseqlevel*\unitfactor-0.7*\unitfactor) node (mess to) {};
	\draw[->,>=angle 60] (mess from) -- (mess to) node[midway, above]
	{#3};
	
	\node (\detokenize{#3} from) at (mess from) {};
	\node (\detokenize{#3} to) at (mess to) {};
}

\newtheorem{protocol}{Protocol}

\newcommand{\hash}{\mathrm{hash}}
\newcommand{\id}{\mathrm{id}}
\newcommand{\removedtodo}[2][]{}

\begin{document}

\title{Binding of Endpoints to Identifiers by On-Chain Proofs\\
\thanks{}
}

\author{\IEEEauthorblockN{Diego Pennino}
\IEEEauthorblockA{\textit{Engineering Dept.} \\
\textit{Roma Tre University}\\
Roma, Italy \\
pennino@ing.uniroma3.it\\
0000-0001-5339-4531}
\and
\IEEEauthorblockN{Maurizio Pizzonia}
\IEEEauthorblockA{\textit{Engineering Dept.} \\
\textit{Roma Tre University}\\
Roma, Italy \\
pizzonia@ing.uniroma3.it\\
0000-0001-8758-3437}
\and
\IEEEauthorblockN{Andrea Vitaletti}
\IEEEauthorblockA{\textit{Engineering Dept.} \\
\textit{Sapienza}\\
Roma, Italy \\
vitaletti@diag.uniroma1.it\\
0000-0003-1074-5068}
\and
\IEEEauthorblockN{Marco Zecchini}
\IEEEauthorblockA{\textit{Engineering Dept.} \\
\textit{Sapienza}\\
Roma, Italy \\
zecchini@diag.uniroma1.it\\
0000-0002-2280-9543}
}

\maketitle              

\begin{abstract}
	
In many applications, identity management (IdM) is used to associate a subject
public key with an \emph{endpoint} at which the subject can be contacted
(telephone number, email, etc.). In decentralized applications based on
distributed ledger technologies (DLTes), it is desirable for the IdM to be
decentralized as well. Currently, endpoints are either verified by who needs it,
which is impractical in DLT-based applications, or by a centralized authority,
which contrasts with the spirit of DLTes.

In this paper, we show two DLT-based protocols to prove the association between
a subject and an endpoint in a decentralized manner, contributing in filling the
gap of the current IdM approaches with respect to decentralization. Our
protocols are compatible with a wide variety of endpoints. We analyze the
security of our protocols and evaluate their performance and cost against
the common approaches.

\end{abstract}

\section{Introduction}

The main purpose of identity management (IdM) is to bind an identifier of a subject (usually a public key) with attributes, claims, entitlements, or properties of that subject~\cite{ISO-24760}. In this paper, we focus on the process to bind a subject identifier with an \emph{endpoint} of a \emph{communication technology} (or \emph{channel}).  In practice, an endpoint identifies a way to contact the subject. Examples of endpoints are web URLs, IP addresses, email addresses, telephone numbers, postal mail addresses,
etc.

Traditionally, IdM has been designed over of centralised architectures and it consequently requires trust in third parties. 
This might be a problem when an IdM is used in applications designed for Distributed Ledger Technologies (DLTs) that are supposed not to rely on any centralized source of trust. As an example, a subject could store in-chain his/her consent to receive phone calls for
marketing purposes. This can be easily implemented storing the consent on-chain  in the form  of a transaction originating from one of the subject public key(s), but there must be some evidence that the subject's public key is bound to his/her phone number.

While for some applications the possession of an endpoint may be a value per
se, in certain countries, certain endpoints are bound by law to a legal person
or a natural person. For example, this is the case for telephone numbers in
Italy. In these cases, the endpoint verification  provides a big added value.

A subject can prove the binding of one of its identifiers with an endpoint by sending or receiving messages on it. This technique is widely used in the context of multi-factor authentication. 
However, currently, its adoption in a decentralized approach requires endpoint verification to be independently performed by each \emph{verifier} that is interested in this information. This might be a problem, if verifiers are many, since each verification is time and resource consuming (see Section~\ref{sec:background}).

\textbf{Contribution of the paper.} In this paper, we present two DLT-based protocols to perform a \emph{decentralized verification process} capable to write in-chain the \emph{proof} (or \emph{certification}) of the binding between a subject identifier and an endpoint, providing a substantial contribution to a  decentralized IdM approach. That proof, can then be used to certify the binding without the burden of the many single verification processes.
We analyze the security of our approach against miscertification
and denial of service attacks. We also analyze the practical applicability of
the proposed protocols with several kinds of endpoints. 

\textbf{Structure of the paper.} In Section~\ref{sec:sota}, we review the state of the art.
In Section~\ref{sec:background} we review basic verification processes and introduce notation.
In Section~\ref{sec:decentralizedCA}, we describe the two protocols to write in-chain the proof of the binding
and, in Section~\ref{sec:security_analysis}, we analyze their security. 
In Section~\ref{sec:evaluation}, we present a first evaluation of our approach and,
in Section~\ref{sec:applications}, we discuss applicability aspects with several kinds of endpoints. 

\section{State of the Art}\label{sec:sota}

IdM is standardized by ISO~\cite{ISO-24760} and there regulations about it (see, for example, \emph{eIDAS}~\cite{eIDAS} a regulation of the European Union). Protocols and standards related to IdM
systems are surveyed in~\cite{10.1093/ietisy/e89-d.1.112}. Identities can be
useful across several organizations. For this reason, \emph{single sign-on}
approaches, such as Facebook connect, are adopted~\cite{10.1007/3-540-45067-X_22}, but usually relays on centralized architectures. The idea of realizing
IdM on top of DLTes is a further step toward making IdM independent from a
specific organization. In private/permissioned DLTes some kind of trust among
participants exists, hence implementing IdM over them does not introduce
new conceptual problems. The \emph{Self-Sovereign Identity (SSI)} approach, surveyed here~\cite{MUHLE201880}, envisions solutions in which subjects should be able to create and control their own identity, without relying on any centralized authority. In this context,
public/permissionless DLTes are fundamental tools. W3C has ongoing efforts to
standardize the building blocks of SSI. \emph{Decentralized Identifiers
	(DIDs)}~\cite{w3cDID1.0} are controlled by subjects and possibly securely stored
in DLTes. DIDs are linked with \emph{DID documents} where attributes are listed.
Certain attributes are associated to \emph{Verifiable Claims/Credentials}
(\emph{VC})~\cite{w3cVCDM1.0} which allow the binding between the identifier and the attributes.
Our approach can be seen as a tool to implement a verifiable claim to prove the binding between a subject identifier and an endpoint

A realization of this framework is backed by the
Decentralised Identity Foundation\footnote{
	\url{https://identity.foundation/}}. The relation between DIDs and eIDAS is analyzed
in~\cite{SSIEIDAS}.

One of the first attempts to design an IdM system deployed on the blockchain
trying to accomplish self-sovereign identities is
Namecoin~\cite{DBLP:conf/weis/KalodnerCEBN15}. The uPort system~\cite{uport}
makes use of Ethereum smart contracts and allows subjects to record simple
statements about them. Sovrin~\cite{sovrin} is an open source identity network
built on a permissioned DLT to manage DIDs in which only trusted institutions
take part in consensus protocols. ShoCard~\cite{shocard} is a digital identity
card for mobile devices. It binds an existing trusted credential (e.g., a
passport),  with additional identity attributes by means of Bitcoin
transactions. The last three systems are also analyzed and compared
in~\cite{8425607}.
Sora~\cite{8377927} and DNS-IdM~\cite{DNS-IdM} are two further recent proposals.

\section{Background and Notation}\label{sec:background} 
Each subject $S$ on a blockchain is associated with at least one pair $\langle
p, s \rangle$, where  $p$ is a public key and $s$ is the corresponding secret
key. We denote by $[m]_s$ or $[m]_S$ the signature of a string $m$ performed by $S$ using
$s$. Each public key of  $S$ can be used as a pseudonym for $S$ when $S$ have to
be mentioned in any transaction recorded in the blockchain. Subject $S$ can
easily prove its association with $p$ by considering a publicly known random
string $r$ (a \emph{challenge}), never used before, whose generation is not
controlled by $S$, and providing $[r]_s$. In interactive protocols with two
parties, $r$ can be generated by the party that needs the proof. In the
blockchain context, $r$ is a cryptographic hash of some \emph{new} piece of
data. For example, $r$ can be the cryptographic hash of the last block or a
string derived by a new transaction to be submitted (usually a transaction to be
valid should be different from all previous ones).

In this paper, we are concerned with a \emph{verifier} $V$ that intends to assess that
a subject $S$ possesses an endpoint $E$. The vast majority of techniques to
achieve this result can be traced back to the following two interactive
protocols that assumes that $V$ already has an alternative communication mean
$M$ with $S$.

\begin{protocol}\label{proto:basic:Sreceives}
\end{protocol}
\begin{enumerate}
	\item $V$ chooses a challenge code $c$ and sends it to endpoint $E$.
	\item $S$ receives $c$ at endpoint $E$ and sends back $c$ to $V$ by $M$.
\end{enumerate}

\begin{protocol}\label{proto:basic:Ssends}
\end{protocol}
\begin{enumerate}
	\item $V$ chooses a challenge code $c$ and sends it to $S$ by $M$.
	\item $S$ sends back $c$ to $V$ from endpoint $E$.
\end{enumerate}

These protocols can be adapted to bind the fact that $S$ possesses $E$
with a key pair $\langle p, s \rangle$ of $S$. In the adapted protocols,
$S$ sends back to $V$ a signed version of $c$ ($[c]_s$), actually
proving that the subject that knows $s$ also possesses $E$. In this
variation, $c$ is essentially a challenge. In the rest of this
paper, we refer to the adapted versions of Protocols~\ref{proto:basic:Sreceives}
and~\ref{proto:basic:Ssends} as \emph{basic protocols}.

In a blockchain context, we potentially have a large number of subjects
and verifiers. We recognize that the basic protocols have the following
drawbacks.
\begin{itemize}

\item Each time a different verifier intends to assess if $S$ possesses $E$, a
new verification should be performed. This might be a serious problem if
verifiers are many and/or verification has a cost for $S$.

\item Each verification requires to set up an interactive protocol. This means
that either $S$ should always be on-line or $V$ should be willing to wait for
$S$ to reply to the challenge.

\end{itemize}

In the spirit of the blockchain, we look for a decentralized solution 
to this certification problem that does not suffer these two drawbacks.

\section{DLT-Based Proofs for Bindings Endpoints to Subjects}\label{sec:decentralizedCA}

A \emph{subject} $S$ intends to obtain a \emph{proof} that (s)he owns an
end-point \ep, to be shown to any interested \emph{verifier}. In this section, we show two blockchain-based
protocols to achieve this goal in a decentralized manner.

We assume that the communication technology of \ep can support the exchange of a message at least 
as large as a (possibly signed) challenge, i.e., a large enough random number. In
Protocol~\ref{proto:subj_to_committee}, the proof-of-possession of \ep is a proof
that $S$ can \emph{send} from \ep a signed
challenge  to a number of other subjects. In Protocol~\ref{proto:committee_to_subj},
vice-versa, the proof-of-possession is a proof that $S$ can
\emph{receive} at \ep  a challenge from a number of
other subjects.

Formally, $S$ intends to obtain a credible proof of a pair $\langle p, \textrm{\ep}\rangle$ meaning that $S$, with public key $p$, owns \ep. We can also write
$\langle S, \textrm{\ep} \rangle$ with the same meaning. When $S$ obtains the proof for
$\langle S,\textrm{\ep} \rangle$, we say that $S$ is \emph{certified}.
For simplicity, in the following, we assume that all subjects that are already
certified are always \emph{on-line}, in the sense that they can
\begin{enumerate}
	
\item interact with \ep by receiving messages at \ep or sending messages from \ep,

\item access the blocks of the blockchain including the last one,

\item monitor the transactions accepted in the blockchain and react to them. 
\end{enumerate} 
We relax some of these assumptions in Section~\ref{sec:correctness_non-ideal}. 
We now introduce our two protocols.

\subsection{Certifying that a Subject Can Send a Message from an Endpoint}

\begin{figure}
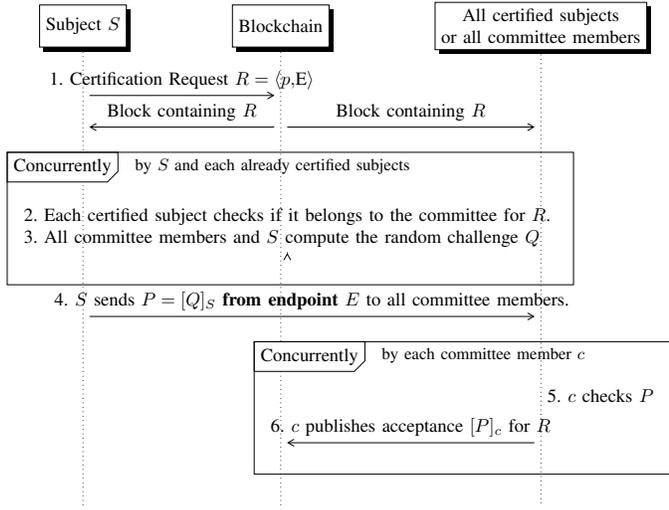

	\begin{center}
		\begin{adjustbox}{scale=0.7}
		\begin{sequencediagram}
			\newinst{S}{Subject $S$}
			\newinst[2]{BC}{Blockchain}
			\newinst[2]{N}{\shortstack{All certified subjects\\ or all committee members}}
			\mess{S}{1. Certification Request $R=\langle p,\ep \rangle$ }{BC}
		    \mess{BC}{ Block containing $R$ }{N}
		    \prelevel\mess{BC}{ Block containing $R$ }{S}
			\begin{sdblock}{ Concurrently }{ by $S$ and each already certified subjects  }
			   \postlevel
			   \begin{mess}{BC}{\shortstack[l]{2. Each certified subject checks if it belongs to the committee for $R$.\\
			   	 3. All committee members and $S$ compute the random challenge $Q$}}{BC}
			    \end{mess}

			\end{sdblock}

		    \mess{S}{4. $S$ sends $P=[Q]_S$ \textbf{from endpoint $E$} to all committee members.}{N}

			\begin{sdblock}{ Concurrently }{ by each committee member $c$ }
			    
			    \node (zzz) at  (mess to)  {};
			    \path (zzz) ++(270:15mm)
			     node[anchor=west]  {\shortstack[l]{5. $c$ checks $P$}};
			    
			    \postlevel
			    \mess{N}{6. $c$ publishes acceptance $[P]_c$ for $R$}{BC}
			    
			\end{sdblock}
			
		\end{sequencediagram}
		\end{adjustbox}
	\end{center}
	
	\caption{ Sequence diagram for Protocol~\ref{proto:subj_to_committee}, where
	$S$ sends signed messages to committee members from endpoint \ep (see Step~\ref{P:step:response_from_E}). }

	\label{fig:proto:subj_to_committee}
\end{figure}

In our frist protocol, subject $S$ with public key $p$ obtains the
proof-of-possession of endpoint \ep by sending signed messages from \ep to a
number of different other subjects. This idea is developed in
Protocol~\ref{proto:subj_to_committee}. The associated diagram is shown in
Figure~\ref{fig:proto:subj_to_committee}. At the end of the protocol, the certificate of possession is 
published in the blockchain.

\begin{protocol}\label{proto:subj_to_committee}
\end{protocol}

\begin{enumerate}

\item \textbf{Certification Request.} Subject $S$ creates \emph{certification
	request} $R=\langle p,\textrm{\ep} \rangle$ (where $p$ is a public key of $S$) and 
publishes $[R]_S$ in the blockchain. Request $R$ reaches all participants as its block is broadcasted.

\item \textbf{Committee Selection.} A committee of $k$ certified subjects, is
randomly selected, on the basis of $R$, following the procedure detailed below.
This is autonomously performed by whoever needs to know the committee
composition, comprising $S$ and all DLT participants. Depending on the communication 
technology, the endpoints of the committee members may be needed, they can be obtained
from their certificates published in the blockchain.

\item \textbf{Random Challenge}. From the block $B$ where $R$ is accepted all
committee members and $S$ computes $Q=\hash(R|B)$.
    
\item \textbf{Response from \ep}. \label{P:step:response_from_E} $S$ sends from \ep to all
committee members the proof $P=[Q]_S$ that $S$ can send a message from 
\ep. 

\item \textbf{Committee checks.} Each committee member $c$  that receives $P$ checks
its signature and value of $Q$.

\item \textbf{Acceptance.} If checks are successful, $c$ publishes on the blockchain its
acceptance $[P]_c$ related to $R$. 

\end{enumerate}

\paragraph{Certificate Verification} A verifier can check the validity of the
acceptance transactions related to $R$ and count them to see if they are $k$, or above
a certain threshold $\bar k \leq k$ (see
Section~\ref{sec:correctness_non-ideal}).

\paragraph{Identifiers and summarization} It is useful to assign to each new certified subject
$S$ a sequential \emph{identifier}, denoted $\id(S)$. The assignment can be done
by a consensus rule to automatically commit a transaction, in a subsequent
block, that states and summarizes the association $\langle S,\textrm{\ep}, \id(S) \rangle$
for each $R=\langle S,\textrm{\ep}\rangle$. This should be done when enough acceptance
transactions for $R$ are committed.

\paragraph{Committee selection procedure}  We assume the network comprises $N$ certified
subjects.  We assume their identifiers to be from
0 to $N-1$, i.e., there are no holes in the sequence of the identifiers. For a 
certification request $R$, we select the committee
$C=\{c_1,\dots,c_k\}$, by selecting the identifiers $\id(c_i)$ of its members.
Suppose that $b$ is the hash of the block in which the certification request $R$
is committed. We use $b$ as a shared source of randomness.

We need a deterministic method that, for each $i \in \{1,\dots,k\}$, randomly selects an
$\id(c_i)\in \{0,\dots,N-1\}$. This method should  depend on $R$ and $b$ and the
selection probability should be uniformly distributed among all already
certified subjects. It is desirable for this association not to select the same
subject twice. However, if $N \gg k$, the probability  of accidentally doubly
selecting the same subject is negligible.

A very simple approach is to select committee members according to the following rule
$$\id(c_i)=\hash(i|R|b) \mod N.$$

Note that, given any deterministic association method, as soon as a
certification request $R$ is published, each certified subject can autonomously
understand if (s)he was selected for the committee to certify $R$. Further,
anyone can easily check if $c_i$ is part of the committee for $R$ as well as
enumerate the whole committee.

\subsection{Certifying that a Subject Can Receive a Message at an Endpoint}

\begin{figure}
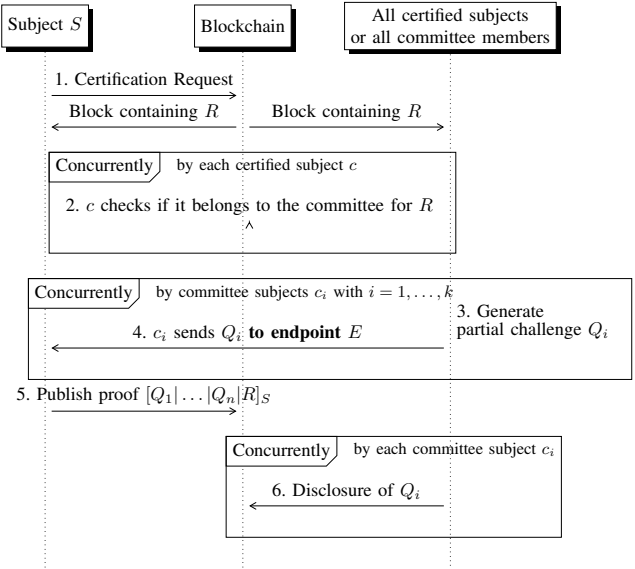

	\begin{center}

		\begin{adjustbox}{scale=0.7}
		\begin{sequencediagram}
			\newinst{S}{Subject $S$}
			\newinst[2]{BC}{Blockchain}
			\newinst[1]{N}{\shortstack{All certified subjects\\ or all committee members}}
			
			\mess{S}{1. Certification Request}{BC}
		    \mess{BC}{ Block containing $R$ }{N}
			\prelevel\mess{BC}{ Block containing $R$ }{S}
			
			\begin{sdblock}{ Concurrently }{ by each certified subject $c$ }
			    
			    \begin{mess}{BC}{2. $c$ checks if it belongs to the committee for $R$}{BC}
			    \end{mess}
			\end{sdblock}

			\begin{sdblock}{ Concurrently }{ by committee subjects $c_i$ with $i=1,\dots,k$ }
			    			    \mess{N}{4. $c_i$ sends $Q_i$ \textbf{to endpoint \ep} }{S}
			    \node (zzz) at  (mess from)  {};
			    \path (zzz) ++(90:5mm)
			     node[anchor=west]  {\shortstack[l]{3. Generate\\ partial challenge $Q_i$}};
			\end{sdblock}

			\mess{S}{5. Publish proof $[Q_1|\dots|Q_n|R]_S$ }{BC}

            \begin{sdblock}{ Concurrently }{ by each committee subject $c_i$  }
			    \mess{N}{6. Disclosure of $Q_i$ }{BC}
			\end{sdblock}

		\end{sequencediagram}
		\end{adjustbox}
		
	\end{center}

	\caption{Sequence diagram for Protocol~\ref{proto:committee_to_subj} in which
	committee members send challenges to endpoint \ep (see Step~\ref{step:challenges_to_E}). }

	\label{fig:proto:committee_to_subj}
\end{figure}

In our second protocol, a subject $S$ obtains the
proof-of-possession of endpoint \ep by receiving challenges at \ep from a 
number of different other subjects. This idea is developed in
Protocol~\ref{proto:committee_to_subj}. The associated diagram is shown in
Figure~\ref{fig:proto:committee_to_subj}. At the end of the protocol, the certificate of possession is 
published in the blockchain.

\begin{protocol}\label{proto:committee_to_subj}
\end{protocol}
\begin{enumerate}
\item \textbf{Certification Request.}
Subject $S$ creates \emph{certification
	request} $R=\langle p, \textrm{\ep} \rangle$ (where $p$ is its public key) and 
publishes $[R]_S$ in the blockchain. Request $R$ reaches all participants as its block is broadcasted.

\item \textbf{Committee Selection.} As in Protocol~\ref{proto:subj_to_committee},
a committee of $k$ members is randomly selected, denoted $C=\{c_1,\dots,c_k\}$.

\item \textbf{Challenges Generation}. Each member $c_i$ of $C$  generates a
random \emph{partial challenge} $Q_i$.

\item \textbf{Challenges to \ep}. \label{step:challenges_to_E} Each $c_i$ sends $Q_i$ to endpoint \ep, where $S$ is supposed to be able to receive it.

\item \textbf{Proof Publication.} \label{step:combine_and_publish_proof} Subject
$S$ combines the partial challenges received to obtain the \emph{complete
	challenge} $Q_1|\dots|Q_k$. $S$ publishes on the blockchain the proof
$P=[Q_1|\dots|Q_k|R]_S$ that $S$ has received all partial challenges and hence that $S$ can
read messages at \ep.

\item \textbf{Challenges Disclosure.} After $P$ is published, each $c_i$ makes 
$Q_i$ public by a suitable transaction, also specifying that it is related to
$R$.

\end{enumerate}

\paragraph{Certificate Verification} A verifier can check the validity of the proof
$P$  and then be sure that the association $\langle p, \textrm{\ep} \rangle$ holds. This
verification can be performed by the following procedure, which takes as input
$P=[Q_1|\dots|Q_k|R]_S$.

\begin{enumerate}

    \item Check that $R=\langle p, \textrm{\ep} \rangle$ is in the blockchain.

    \item Check that $P$ is correctly signed by public key $p$.

    \item For each $i=1,\dots,k$  
    \begin{itemize}
        \item In the blockchain, look for the  transaction containing the challenge disclosure for $Q_i$ related to $R$ performed by $c_i$.

        \item Check that the challenge  disclosure for $Q_i$ appears after $P$ in the history.

    \end{itemize}

    \item If all the previous checks are successful, the proof $P$ is valid and
the association $\langle S, \textrm{\ep} \rangle$ is \emph{certified}.
\end{enumerate}

\paragraph{Identifiers and summarization} As for Protocol~\ref{proto:subj_to_committee}, 
a sequential identifier $\id(S)$ should be assigned to each newly certified subject $S$. 
The assignment can be done
by a consensus rule in which DLT nodes should execute the above verification process and, if successful, 
commit a transaction that states the association $\langle S,\textrm{\ep}, \id(S) \rangle$.

\paragraph{Committee selection procedure} For the committee selection, the same
approach of Protocol~\ref{proto:subj_to_committee} can be adopted.

\subsection{Correctness Under Non-Ideal Conditions}\label{sec:correctness_non-ideal}

Correctness of Protocols~\ref{proto:subj_to_committee} and~\ref{proto:committee_to_subj}
depends on the reliability of the challenge-response process. In particular,
the committee members, the endpoint and the communication technology should be reliable. To tolerate a
margin of committee subjects that are not on-line (or problems during
communication), we can slightly change the protocols.

Protocol~\ref{proto:subj_to_committee} can be easily adapted to tolerate a number of
missing acceptance transactions by just fixing a suitable threshold $\bar k \leq
k$ of members that have to accept the proof, for it to be valid.

Protocol~\ref{proto:committee_to_subj} can also be adapted. In
Step~\ref{step:combine_and_publish_proof}, $S$ may not wait for all partial
challenges but only for $\bar k \leq k$ of them to tolerate a fraction of
missing ones.

A further aspect is that there is no reason for committee members to execute
the certification protocol. To solve this problem, they should be rewarded for their
work. This is not hard to do if the underlying DLT also implements a cryptocurrency, 
which is often the case.

\section{Security}\label{sec:security_analysis}
In this section, we introduce our threat model and analyze  the security of Protocols~\ref{proto:subj_to_committee} and~\ref{proto:committee_to_subj} in that model.

\subsection{Threat Model}

In analyzing the security of our protocols, we consider the following threats (or malicious objectives).

\begin{description}[leftmargin=11em,style=nextline]
    \item[Miscertification.] Certification of untrue association of subject $S$ with endpoint \ep.
    \item[Denial of service (DoS).] Denial of certification for a legitimate $\langle S, \textrm{\ep} \rangle$ association.
\end{description}

We denote by $N$ the number of certified subjects in the network.
In our model, any adversary-controlled (certified or non-certified) subject $S$ may deviate from 
the protocol in an arbitrary way.

We assume the adversary may ask certification for several pairs $\langle S_i, E_i
\rangle$. This is legitimate and easy to achieve, provided that the adversary
actually controls the $E_i$'s and owns keys for subjects $S_i$'s. These
certified subjects may deviate form the protocol and be leveraged to perform 
malicious actions realizing a \emph{Sybil attack} in our context. However, we
assume that only a limited number of subjects $m$, with $m \leq N$, can be certified by the
adversary. This can be enforced in practice in several ways. For example, if the
blockchain supports a cryptocurrency, we may regularly charge each certified subject a fee to keep old certificates active. Hence, we can assume that the cost of controlling $m$ subjects is $c m$ where $c$ is the cost for each subject in a certain period of time.

We assume the underlying blockchain technology to be secure. Hence, the adversary \begin{enumerate}
    \item cannot tamper with the blockchain,
    \item cannot stop a non-controlled subject from asking to the blockchain to perform a transaction, and
    \item cannot stop a non-controlled subject from
reading from the blockchain the committed transactions.
\end{enumerate}
This also means that the adversary cannot add certifications, delete or tamper with past certificates, or change subject identifiers.

We assume the communication technology of \ep to be immune from certain specific
attacks. For both protocols, we assume no network-level denial of service is
possible. For Protocol~\ref{proto:subj_to_committee}, we assume that the
adversary cannot create spoofed messages that look as if they are sent from
\ep. For Protocol~\ref{proto:committee_to_subj}, we assume the adversary cannot eavesdrop
messages sent to \ep.

\subsection{Security Analysis}

To prove the security of our approach against \emph{miscertification}, we should
prove that it is hard for an adversary $A$ to certify an association
$\langle S, \textrm{\ep} \rangle$, when \ep is not actually controlled by $A$. 
We show that the adversary is going to spend $O(N)$ to perform the attack and hence that
the security of our approach increases for larger $N$.

Since the adversary cannot spoof/eavesdrop messages from/to \ep, 
if \ep is not
controlled by $A$, the only way that $A$ has to obtain the certificate is to
corrupt a committee controlling at least $\bar k$ committee members. The corrupted committee members can assert that 
they received or sent the messages through \ep,  as stated by our protocols, even if communication occurred using another channel. 
A committee is formed by
$k$ randomly extracted members, out of $N$ total certified subjects, of whose only $m$ are corrupted. The
probability $p(m)$ that, in a committee, at least $\bar k = \alpha k$ members (with $0<\alpha<1$)  are controlled by $A$ is given by 
$p(m)=\frac{\sum_{i=\bar k}^k {m \choose i} {N-m \choose k-i}  }{{N \choose k}}$.
Function $p(m)$ equals $1/2$ at $m=\alpha N$ and it is very close to zero for $m<3/4 \alpha N$
This means that
when $N$ increases, the adversary have to corrupt at least about $\alpha N$
subjects to have a reasonable probability to get a corrupted committee. 
Hence, for $k$ and $\bar k$ fixed, the adversary should
spend $O(N)$ to perform a miscertification attack with reasonable probability.

To perform a denial of service, $A$ cannot act at the
network level or at the blockchain level (by threat model). The only way for $A$
to block a certification process is to force honest members to be less than $\bar k$. To do this, 
$A$ should corrupt more than $k-\bar k = k
(1-\alpha)$ members.
Hence, the same argument we showed for miscertification applies where we substitute $\alpha$ with $1-\alpha$.

\begin{table*}[t]
	\footnotesize 
	\begin{tabular}{|l|c|M{2cm}|M{2cm}|M{2.5cm}|M{1.5cm}|}
		\hline
		\multirow{2}{*}{\textbf{Endpoint}}
	    & \multirow{2}{*}{\textbf{Message}}
		& \multirow{2}{2cm}{\centering\textbf{Cost to send \bf one message}}
		& \multicolumn{2}{c|}{\textbf{Security (Attack difficulty)}}  
		& \multirow{2}{1.5cm}{\centering \textbf{Suggested Protocol}} \\ 
		\cline{4-5}
		& & & \textit{spoofing (P\ref{proto:subj_to_committee})} 
			& \textit{ eavesdropping (P\ref{proto:committee_to_subj})} &  \\
		\hline
		Phone number & SMS &  SMS cost & \multirow{2}{2cm}{High: \scriptsize telecom op. countermeasures} & \multirow{6}{2.8cm}{Medium: \scriptsize it requires physical or logical proximity to the endpoint to miscertify } & \multirow{2}{1.5cm}{P\ref{proto:subj_to_committee}: \scriptsize to charge cost on subject}  \\
		\cline{1-3}
		Phone number & phone call with IVR & phone call cost &   &   & \\
		\cline{1-4}\cline{6-6}
		Postal mail address & letter& stamp &  Low &   & \multirow{3}{1.5cm}{P\ref{proto:committee_to_subj}: \scriptsize for security reason} \\
		\cline{1-4}
		Email address & email & negligible  &  Low &  &  \\
		\cline{1-4}
		IP address & IP packet &negligible &  Low &   &  \\
		\cline{1-4}\cline{6-6}
		Web page & \shortstack{Page change and HTTP response for P\ref{proto:subj_to_committee} \\ HTTP GET/POST for P\ref{proto:committee_to_subj} {\scriptsize (technically complex)}}&negligible &  \shortstack{Medium: \\{\scriptsize proximity required}}  &    & P\ref{proto:subj_to_committee}: {\scriptsize easier to implement}\\
		\cline{1-4}\cline{5-6}
		DNS name & DNS response (P3 only) & negligible &  High: {\scriptsize distributed} &  -  & P\ref{proto:subj_to_committee} \\
		\hline 
		\shortstack[l]{Bank account/IBAN } & \emph{description} field of bank transfer  & transfer cost &  High & High& P\ref{proto:subj_to_committee} or P\ref{proto:committee_to_subj}\\
		\hline
			\end{tabular}

	\caption{Summary of endpoint features. For brevity, Protocols~\ref{proto:subj_to_committee} and~\ref{proto:committee_to_subj} are referred as P\ref{proto:subj_to_committee} and P\ref{proto:committee_to_subj}. }
	\label{table:appcontext}
\end{table*}

\section{ Performance Evaluation}\label{sec:evaluation}

In this section we provide a comparison analysis between the basic protocols
(see Section~\ref{sec:background}) and the decentralized ones (see
Section~\ref{sec:decentralizedCA}) in terms of latency and of messages
transmitted through the endpoint.

We start by analyzing the time taken by decentralized protocols from when the transaction with the 
certification request $R$ is broadcasted to when the certification is known to all nodes. We
call this interval of time the \emph{latency} of the protocol. We assume the delay for
creating and signing a challenge to be negligible with respect to other
delays involved in the protocol. 

\begin{figure}
	\centering
	\includegraphics[width=0.7\linewidth]{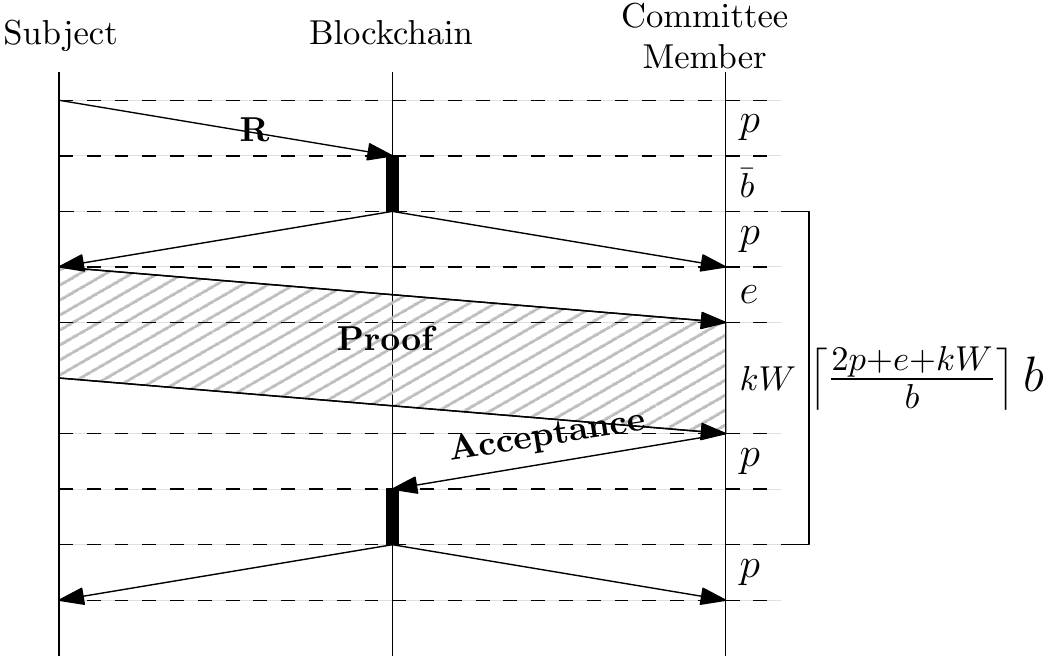}
	\caption{Timings for Protocol~\ref{proto:subj_to_committee}.}
	\label{fig:protocol3latency}
\end{figure}
\begin{figure}
	\centering
	\includegraphics[width=0.7\linewidth]{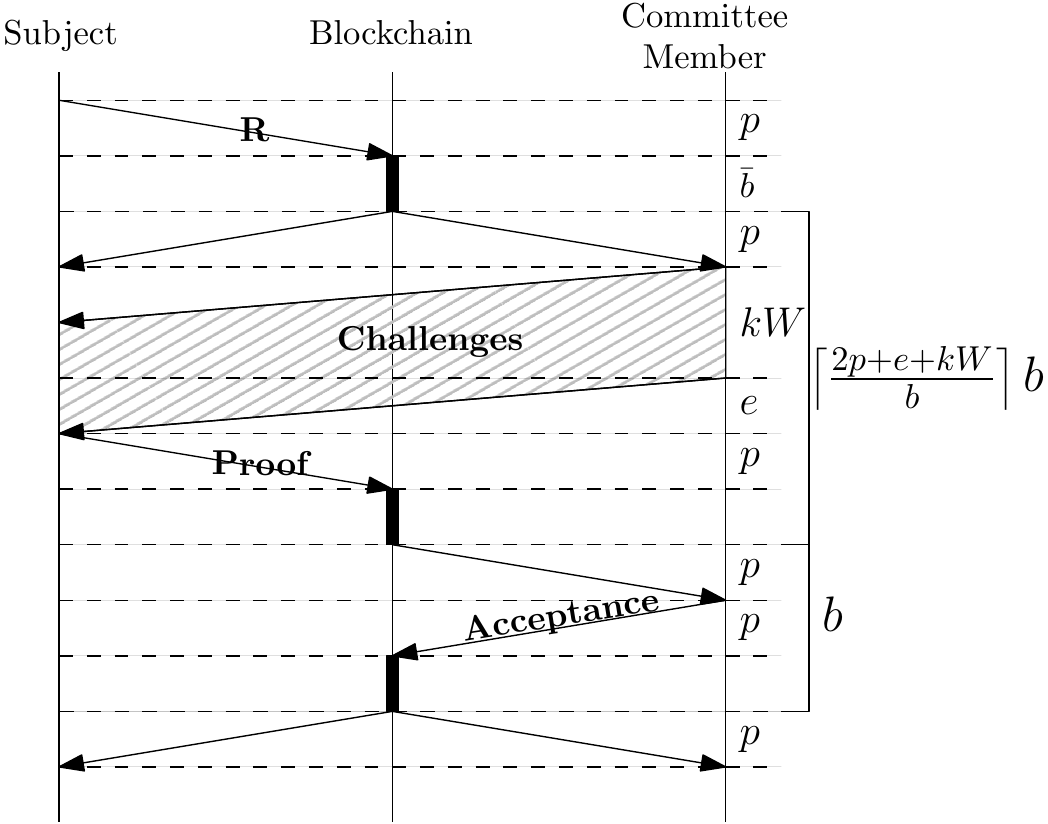}
	\caption{Timings for Protocol~\ref{proto:committee_to_subj}.}
	\label{fig:protocol4latency}
\end{figure}

We call $b$ the interval of time between two consecutive blocks. For simplicity, we assume $b$ to be constant and
the number of transactions a single block can host is greater than the committee size $k$. We
denote by $\bar b<b$ the time that a transaction waits before acceptance.  The expected value of $\bar b$ is $b/2$, 
if the instant in which a new transaction is received by the blockchain is independent from previous block commit time. We call $p$, with $p\ll b$, the time to propagate a block to all
nodes. We assume that the time to propagate a new (still unaccepted) transaction to all nodes
is also $p$. We call $W$ the time that a single message takes to be transmitted
to the endpoint and $e$ the time it takes to arrive to destination, hence, $k$ messages
are received after $e+kW$ time.

Latency of Protocol~\ref{proto:subj_to_committee} is easily derived by observing the
diagram in Figure~\ref{fig:protocol3latency} and it turns out to be $\left\lceil
\frac{2p+e+kW}{b} \right\rceil b+2p+\bar b$. Analogously by observing the
diagram in Figure~\ref{fig:protocol4latency}, we derive the latency of
Protocol~\ref{proto:committee_to_subj}, which is given by $\left(\left\lceil
\frac{2p+e+kW}{b} \right\rceil+1 \right)b+2p+\bar b$.  We note that, for both protocols, after
certification is recorded in the chain, a verifier takes negligible time to
check it.

About the latency of basic protocols, let $v$ be the number of verifiers. We assume each verifier verifies $\langle S, \textrm{\ep} \rangle$ only one time.
We assume that the alternative communication mean $M$ between the subject and each verifier has negligible
latency and transmission time.  The latency of the basic protocols turns out to
be $e+vW$.

Concerning the number of messages that have to be transmitted through the
endpoint, the basic protocols send $v$ messages overall, while both our
protocols send only $k$ messages.

Since our protocols use the endpoint  only to transmit a constant number of
messages during the certification, they turn out to be advantageous when
$v > k$. We note that $k$ can be tuned to obtain a spectrum of tradeoffs between 
efficiency and security. 
We also note that our protocols turns out to be advantageous also with
respect to latency when $W$ is large and 
$v \gg k$.

\section{Application Contexts}\label{sec:applications} Nowadays, users 
already provide proofs-of-possession regarding endpoints in a number of
situations. Since, they can be easily turned into use cases for our protocols,
we briefly review some of them in this section. Table~\ref{table:appcontext}
summarizes the most common cases. The table reports the endpoint, the kind of
message used to proof its possession, the cost, a brief remark about the
security regarding its use in our protocols, and the suggested protocol choice.
 Phone numbers can be certified by both SMS or IVR-based phone
calls. Regular postal addresses can be certified by sending letters, even if they may be impractical
for large committees.
A static web page is easy to use as a broadcast channel in the direction from
server to browser(s) for Protocol~\ref{proto:subj_to_committee}. The opposite
direction is also an option, but it requires the web site owner to be able to
get details of the GET/POST request. DNS records can also be used as broadcast
channel for Protocol~\ref{proto:subj_to_committee} while the opposite direction
is not available since queries are usually served by third party servers.
For bank transfers, a small amount to
transfer might also be needed. Concerning security, in most cases eavesdropping
is easy if the attacker can reach a position that is close to the endpoint that
(s)he intends to miscertify. This is true also for SMS (which are weekly
encrypted), but this risk is usually accepted in two-factor authentication
procedures. In certain cases, it might be possible to mitigate this risk by
using encryption (e.g, as for SSL/TLS for IP), but this may increase complexity
and number of messages substantially. For the web, the spoofing attack requires
first to eavesdrop the request, hence proximity to endpoint is required. For DNS,
responses are generated by third party servers in a distributed manner, hence,
the attacker cannot exploit endpoint proximity. Eavesdropping on the committee side is 
hard, as well. Concerning bank transfers, we assume that confidentiality and integrity are assured by
the interbank communication network.

\section{Discussion and Conclusions}

We described two blockchain-based decentralized protocols to create verifiable
claims regarding ownership of a certain endpoint. We believe that our
contribution complements the use of decentralized identifier and is a further
step toward a fully decentralized IdM process.

With respect to currently adopted naive approaches, our protocols 
do not require the subject to be on-line and the endpoint load
does not depends on the number of verifications.
This makes our approach especially suited in applications where 
a large number of subjects that are not always on-line
 are needed to be verified by a large number of verifiers.


\begin{thebibliography}{10}

\bibitem{ISO-24760}
Iso/iec 24760:2019 it security and privacy — a framework for identity
  management.
\newblock Technical report, New York, 1994.

\bibitem{eIDAS}
European Union.
\newblock {eIDAS} -- {Electronic Identification, Authentication and Trust
  Services}, {EU Regulation} 910/2014.

\bibitem{10.1093/ietisy/e89-d.1.112}
Teruko Miyata, Yuzo Koga, Paul Madsen, Shin-Ichi Adachi, Yoshitsugu Tsuchiya,
  Yasuhisa Sakamoto, and Kenji Takahashi.
\newblock A survey on identity management protocols and standards.
\newblock {\em IEICE - Trans. Inf. Syst.}, E89-D(1):112–123, January 2006.

\bibitem{10.1007/3-540-45067-X_22}
Andreas Pashalidis and Chris~J. Mitchell.
\newblock A taxonomy of single sign-on systems.
\newblock In Rei Safavi-Naini and Jennifer Seberry, editors, {\em Information
  Security and Privacy}, pages 249--264, Berlin, Heidelberg, 2003. Springer
  Berlin Heidelberg.

\bibitem{MUHLE201880}
Alexander Mühle, Andreas Grüner, Tatiana Gayvoronskaya, and Christoph Meinel.
\newblock A survey on essential components of a self-sovereign identity.
\newblock {\em Computer Science Review}, 30:80 -- 86, 2018.

\bibitem{w3cDID1.0}
D~Reed and M~Sporny.
\newblock {W3C} {Decentralized Identifiers} ({DIDs}) v1.0, 2017.

\bibitem{w3cVCDM1.0}
M~Sporny, D~Longley, and Chadwick D.
\newblock {W3C} {Verifiable Credentials Data Model} 1.0: Expressing verifiable
  information on the web, 2018.

\bibitem{SSIEIDAS}
EU.
\newblock {eIDAS} supported self-sovereign identity.
\newblock
  \url{https://ec.europa.eu/futurium/en/system/files/ged/eidas_supported_ssi_may_2019_0.pdf},
  Last accessed April 2020.

\bibitem{DBLP:conf/weis/KalodnerCEBN15}
Harry~A. Kalodner, Miles Carlsten, Paul Ellenbogen, Joseph Bonneau, and Arvind
  Narayanan.
\newblock An empirical study of namecoin and lessons for decentralized
  namespace design.
\newblock In {\em 14th Annual Workshop on the Economics of Information
  Security, {WEIS} 2015, Delft, The Netherlands, 22-23 June, 2015}, 2015.

\bibitem{uport}
{uPort}: a self-sovereign identity and user-centric data platform.
\newblock \url{https://github.com/uport-project/specs}, Last accessed April
  2020.

\bibitem{sovrin}
sovrin.
\newblock sovrin: Control your digital identity.
\newblock Last accessed April 2020.

\bibitem{shocard}
ShoCard.
\newblock Shocard whitepaper.
\newblock Last accessed April 2020.

\bibitem{8425607}
P.~Dunphy and F.~P. Petitcolas.
\newblock A first look at identity management schemes on the blockchain.
\newblock {\em IEEE Security \& Privacy}, 16(04):20--29, jul 2018.

\bibitem{8377927}
M.~{Takemiya} and B.~{Vanieiev}.
\newblock Sora identity: Secure, digital identity on the blockchain.
\newblock In {\em 2018 IEEE 42nd Annual Computer Software and Applications
  Conference (COMPSAC)}, volume~02, pages 582--587, 2018.

\bibitem{DNS-IdM}
Jamila Kassem, Sarwar Sayeed, Hector Marco-Gisbert, Zeeshan Pervez, and Keshav
  Dahal.
\newblock Dns-idm: A blockchain identity management system to secure personal
  data sharing in a network.
\newblock {\em Applied Sciences}, 9:2953, 07 2019.

\end{thebibliography}
\end{document}